\documentclass[runningheads]{llncs}
\usepackage[T1]{fontenc}
\usepackage{graphicx}
\usepackage[colorlinks=true, linkcolor=blue, citecolor=blue, urlcolor=blue]{hyperref} 
\usepackage{tikz}
\usepackage{pgf-pie}
\usepackage{wrapfig} 
\begin{document}
\title{Scaling Accessibility Education: Reflections from a Workshop Targeting CS Educators and Software Professionals}
\titlerunning{Reflections from an Accessibility Workshop}

\author{P D Parthasarathy \orcidID{0000-0002-8723-2407} \and
Anshu M Mittal \orcidID{0009-0004-4481-8645} \and
Swaroop Joshi \orcidID{0000-0003-4536-2446}}

\authorrunning{Parthasarathy et al.}

\institute{BITS Pilani, KK Birla Goa Campus, Goa, India \\
\email{\{p20210042, f20231125, swaroopj\}@goa.bits-pilani.ac.in}\\}

\maketitle              % typeset the header of the contribution
\begin{abstract}
Despite growing global attention to digital accessibility, research from India highlights a significant gap in accessibility training for both computing educators and software professionals. To address this need, we designed and conducted an experiential workshop aimed at building foundational capacity in accessibility practices among 77 participants, including computer science (CS) faculty and industry practitioners. The one-day workshop combined hands-on activities, tool demonstrations, and case studies to foster practical understanding and engagement. Post-workshop feedback showed that a majority of participants rated the workshop positively, with many reporting increased confidence and a shift in their perception of accessibility as a shared responsibility. Additionally, participants expressed a strong interest in applying accessibility principles within their workplaces, underscoring the workshop’s practical relevance and impact. In this \emph{experience report}, we detail the workshop’s design, implementation, and evaluation, and offer actionable insights to guide future initiatives aimed at strengthening accessibility capacity across India’s computing education and professional landscape.

\end{abstract}

\section{Introduction}

Digital accessibility refers to the inclusive practice of designing and developing digital technologies, platforms, and content to ensure that they are usable and accessible to all individuals, including persons with disabilities. It aims to eliminate barriers that prevent people with diverse abilities—such as visual, auditory, cognitive, motor, or speech impairments—from effectively engaging with digital environments. The core objective is to guarantee equal participation in the digital world, enabling individuals with disabilities to access and benefit from digital content and services on par with others. Far from being a niche concern, digital accessibility is a foundational requirement for all digital artifacts and plays a pivotal role in fostering an inclusive digital society.

To support the creation of accessible digital products, the World Wide Web Consortium (W3C) introduced the Web Content Accessibility Guidelines (WCAG) in 1999, most recently updated in version 2.2~\cite{w3c_web_2024}. WCAG is structured around four guiding principles—Perceivable, Operable, Understandable, and Robust (POUR)—and defines three levels of conformance: (a) Level A: the minimum baseline for accessibility, (b) Level AA: a commonly targeted intermediate standard, (c) Level AAA: the highest and most comprehensive standard.

Importantly, accessibility\footnote{Throughout this paper, `accessibility' refers to digital accessibility.} is not optional; it is mandated by law in many countries, including the United States~\cite{the_us_government_it_2023}, the European Union~\cite{europian_commission_web_2023}, Canada~\cite{secretariat_standard_2011}, and India, under the Rights of Persons with Disabilities (RPwD) Act, 2016~\cite{ministry_of_social_justice_and_empowerment_government_of_india_rights_2016}. Despite such legal frameworks, global compliance remains critically low. According to the 2025 WebAIM report~\cite{webaimWebAIMMillion2024}, only 5.2\% of the top one million websites globally meet WCAG Level AA standards. Even after 25 years of WCAG, most websites still fail to meet even Level A compliance~\cite{branham_state_2024}, with only a 3.1\% decline in Level A failures over the past six years~\cite{webaimWebAIMMillion2024}.

Digital inaccessibility also carries significant economic consequences. According to the 2021 UserWay report~\cite{userway_economic_2021}, the global e-commerce sector loses an estimated \$16 billion annually due to inaccessible websites and apps. A global survey of 107 industry leaders found that only 3\% felt their workforce had the skills to meet accessibility goals, and just 2\% found it easy to hire professionals with such expertise~\cite{teach_access_accessible_2023}. A growing body of research identifies the lack of awareness and insufficient accessibility skills among developers as major contributors to inaccessible digital products~\cite{patel_why_2020,parthasarathy_exploring_2024,lazar_improving_2004,webaim_survey_2021,bi_accessibility_2022,parthasarathy_skill_2025}.

A key reason for this gap is the limited integration of accessibility in Computer Science (CS) curricula~\cite{patel_why_2020,parthasarathy_exploring_2024,parthasarathy_teaching_2024,shinohara_who_2018,soares_guedes_how_2020}. While several accessibility education initiatives have been documented in North America and Europe, there is a striking lack of such efforts in the Global South, particularly in India, a country that is home to the world’s largest population of computing students and software professionals. As India continues to lead the global technology workforce, equipping its developers, students, and educators with accessibility knowledge is both a social imperative and an economic necessity. Research from India highlights an urgent need for accessibility training among both educators and professionals~\cite{parthasarathy_teaching_2024,parthasarathy_exploring_2024}. To address this gap, we designed and delivered a workshop for software professionals and CS educators. In this article, we present our experiences in designing, implementing, and evaluating the workshop, along with key insights and lessons learned. Our goal is to contribute actionable knowledge toward building accessibility capacity in India's computing education and professional ecosystem.

In the remainder of this paper, we first provide a brief review of relevant literature (Sec.\ref{sec:RelatedWork}). We then describe the workshop in detail, including the instruments employed to evaluate their effectiveness (Sec.\ref{sec:workshops}). Next, we present the findings of our study (Sec.\ref{sec:StudyResults}) and discuss key takeaways (Sec.\ref{sec:takeaway}) and conclude in Sec.~\ref{sec:conc}.
\section{Background and Literature Review}
\label{sec:RelatedWork}

\subsection{Accessibility Skills Gap and the Need for Accessibility Training}
Numerous studies indicate that formal education often fails to adequately equip professionals with the necessary accessibility competencies~\cite{the_partnership_on_employment__accessible_technology_peat_accessible_2018,parthasarathy_exploring_2024,coverdale_digital_2024,lewthwaite_researching_2023}. While accessibility education at the university level is gradually expanding (as discussed in the next section), the training of practicing software engineers within the Information Technology (IT) sector and CS educators remains relatively underexamined in the existing literature.

In 2020, Patel et al.~\cite{patel_why_2020} surveyed 77 technology professionals in the United States to explore their perspectives and challenges related to accessibility. The study identified significant gaps in formal accessibility expertise, a lack of accessible tools and resources, and insufficient consideration of retroactive accessibility adjustments within project timelines. A similar study conducted in India in 2023 with 269 participants~\cite{parthasarathy_exploring_2024} echoed these findings, emphasizing the lack of formal accessibility expertise and inadequate tools and resources among professionals. WebAIM's 2021 global survey of 758 web accessibility practitioners~\cite{webaim_survey_2021} revealed that nearly half of the respondents admitted their organization's web products were `not highly accessible'. Furthermore, 40.6\% reported no significant improvement in accessibility compared to previous years, while 13\% noted a decline in accessibility standards. In addition, a 2022 study by Bi et al. ~\cite{bi_accessibility_2022}, which surveyed 365 professionals across 26 countries and conducted 15 follow-up interviews, revealed a widespread lack of accessibility expertise and highlighted the critical need for both executive-level support and targeted technical training.

%Crabb et al.~\cite{crabb_developing_2019} conducted an accessibility design workshop involving 197 software developers to explore their understanding and implementation of accessible design practices. The study revealed notable gaps in participants’ knowledge of accessibility, despite existing efforts to promote inclusive design principles. 
A 2023 report by Teach Access, based on insights from 107 tech leaders, revealed a significant accessibility skills gap: 56\% of organizations struggle to find candidates with accessibility expertise, while only 2\% find it easy to do so. Furthermore, 44\% of respondents reported that their current staff lacked adequate accessibility skills, with 75\% noting an increased demand for these skills over the past five years and 86\% anticipating further growth in this demand~\cite{teach_access_accessible_2023}.

%These findings collectively highlight a persistent gap in accessibility awareness and its practical implementation. Addressing this gap requires not only increased training opportunities but also systemic support to integrate accessibility as a core competency in professional and academic settings.

\subsection{Accessibility Education}
To address the accessibility skills gap, numerous studies—primarily from Western contexts—have documented efforts to teach accessibility\footnote{It is important to distinguish between teaching accessibility and teaching accessibly. The former involves instructing students on accessibility philosophy, principles, and tech such as WCAG, while the latter focuses on ensuring that course materials, learning management systems, and resources are accessible to all learners. The scope of this section and our work is limited to the former.}. In college-level computing courses such as web-designing~\cite{kawas_teaching_2019}, Software Engineering~\cite{el-glaly_teaching_2020,aljedaani_enhancing_2025,aljedaani_accessibility_2025,aljedaani_sprint_2025}, CS1/CS2~\cite{jia_infusing_2021}, and AI~\cite{tseng_exploration_2022}. Parallel efforts in India have also begun to emerge, with accessibility being integrated into courses like mobile application development~\cite{bhatia_integrating_2023} and operating systems~\cite{parthasarathy_reflections_2024}. A systematic review of over 50 papers on accessibility education in computer science identifies four key learning objectives: (a) Accessibility Awareness – understanding abilities, legal frameworks, and ethical considerations; (b) Technical Knowledge – familiarity with guidelines, implementation of accessibility features, and testing methods; (c) Empathy; and (d) Career Opportunities in accessibility fields~\cite{baker_systematic_2020}.

Elglagy et al. \cite{elglaly_beyond_2024} argue for establishing accessibility as its own Knowledge Area within the CS curriculum, presenting structured knowledge units and learning objectives. It is also noteworthy that accessibility and inclusive design have received increased emphasis in the ACM/IEEE/AAAI Computer Science Curricula 2023~\cite{kumar_computer_2024}, where they are designated a separate Knowledge Unit with several accessibility topics listed as \textsc{CS Core}\footnote{As per the ~\cite{kumar_computer_2024}, CS Core topics are topics that every computer science graduate \textbf{must} know.}. In 2024, to support educators in integrating accessibility into core CS courses, a comprehensive, free online textbook has been published~\cite{oleson_teaching_2024}.

%More recently, in 2024, Coverdale et al. investigated how the context of education and training shapes teaching and learning in both university and workplace settings ~\cite{coverdale_digital_2024}. Their analysis highlights that faculty and workplace cultures frequently perpetuate the precarious state of accessibility education by individualizing responsibility, often relegating it to designated ‘heroes’ or ‘champions.’ This approach, combined with disciplinary and role-based silos, limits opportunities for raising awareness and developing widespread accessibility competency. In contrast, centers of excellence and communities of practice (CoPs) have shown promise in bridging the gap between education and research, engaging expert users, and fostering interdisciplinary and cross-role learning environments where accessibility is increasingly recognized as a shared responsibility~\cite{parthasarathy_teaching_2024}. 

Despite growing global interest in accessibility education, there remains a significant gap in both research and practice from the Global South—particularly India, which hosts the largest population of computing students and one of the world’s largest software engineering workforces~\cite{parthasarathy_exploring_2024,parthasarathy_teaching_2024}. In the United States, initiatives like \textit{Teach Access}\footnote{\url{https://www.teachaccess.org/initiatives/}} have played a critical role in building accessibility capacity by bridging the gap between academia and industry and supporting faculty in integrating accessibility into computer science curricula. However, India currently lacks a comparable cross-sector organization or initiative focused on scaling accessibility education across educational institutions and the tech workforce. In response to this gap, we designed and conducted a digital accessibility workshop, targeting a mixed group of software professionals and CS educators. We aimed at exploring how accessibility training can be effectively contextualized, delivered, and assessed within the Indian socio-technical environment, offering practical insights into capacity building across both academic and industry settings. To the best of our knowledge, this is the first report on informal accessibility education efforts of this nature in India. \vspace{-0.35cm}

%While numerous studies have explored accessibility training within Western academic and corporate contexts, limited attention has been given to how accessibility is taught—or neglected—within Indian computing education and professional development. Existing literature highlights a notable absence of formal accessibility instruction in Indian universities, along with low awareness among software professionals and CS educators, underscoring the urgent need for targeted accessibility training~\cite{parthasarathy_exploring_2024,parthasarathy_teaching_2024}.
\section{The Workshop}
\label{sec:workshops}

The workshop was specifically designed for software professionals and CS educators in India. The primary objective of the workshop was to support capacity building in digital accessibility among software professionals and CS educators in India—a group for whom accessibility training opportunities remain limited. The workshop aimed to introduce participants to the fundamentals of accessibility, including both technical concepts and social dimensions that are often overlooked in Indian professional settings. Specifically, the learning objectives of the workshop were:
\begin{itemize}
    \item Participants will be able to recognize and apply principles of disability etiquette and inclusive communication, addressing a critical gap in cultural and professional understanding within the Indian context.
    \item Participants will be able to explain foundational concepts of digital accessibility, including key principles, relevant legal frameworks, and the Web Content Accessibility Guidelines (WCAG).
    \item Participants will be able to use practical techniques and tools for improving accessibility in their work environments, such as implementing accessible design practices, conducting basic accessibility assessments, and integrating accessibility into software development workflows and educational curricula.
    \item Participants will be able to adopt a mindset that frames accessibility as a shared, cross-functional responsibility, spanning roles such as developers, designers, educators, and project managers.
\end{itemize}

To recruit participants, an open invitation was shared widely through multiple channels, including professional social media platforms and academic mailing lists. Additionally, the invitation was personally circulated through the authors' professional networks in both academia and the IT industry to ensure broader reach and relevance. 

\begin{wrapfigure}{r}{0.5\textwidth}
\includegraphics[width=\linewidth]{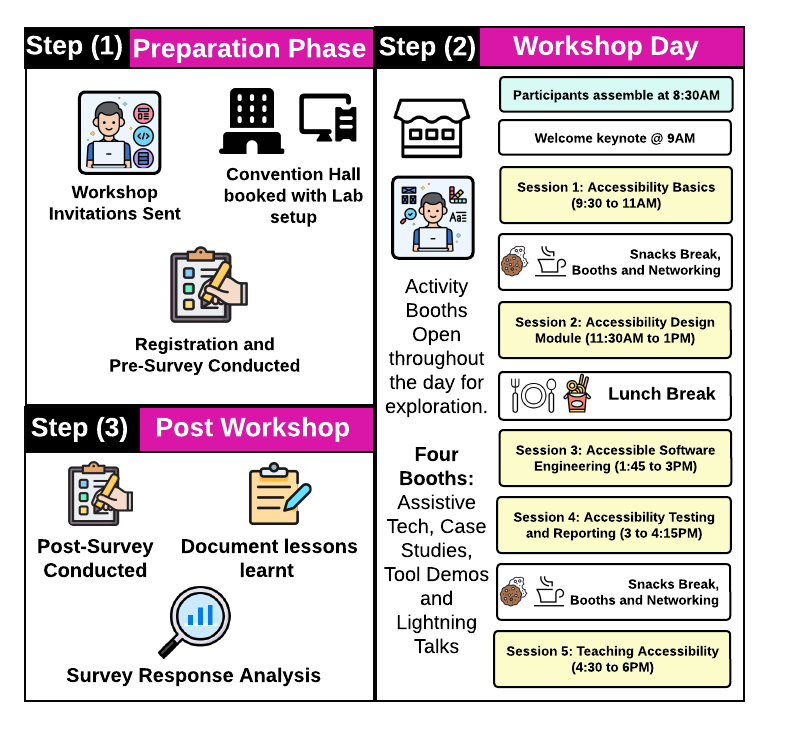}  
\caption{Detailed process of the Workshop}
\label{fig:conf} \vspace{-0.75cm}
\end{wrapfigure}

As shown in Fig. \ref{fig:conf}, the workshop followed a structured three-phase model: preparation, execution, and post-workshop reflection. The one-day event, held around Global Accessibility Awareness Day (GAAD), took place at a dedicated convention hall equipped for interactive, hands-on learning. To help cover logistical expenses such as venue setup and learning materials, a nominal registration fee of 500 INR was collected. The workshop was intentionally designed to be inclusive and beginner-friendly, with no prerequisites for participation—lowering barriers and welcoming a diverse group of attendees from both academia and industry. The day began with participant assembly at 8:30 AM, followed by a keynote address at 9:00 AM by a prominent expert in the accessibility domain. The workshop featured five instructional sessions led by the first author, distributed throughout the day: (a) Accessibility Basics, (b) Accessibility Design Module, (c) Accessible Software Engineering, (d) Accessibility Testing and Reporting, and (e) Teaching Accessibility / Accessibility Enablement. The workshop adopted an active learning approach, where the instructor introduced key concepts and then engaged participants through interactive activities such as think-pair-share and hands-on exploration. Experiential learning was emphasized by encouraging participants to interact with assistive technologies and accessibility testing tools directly on their own laptops. Each instructional module lasted approximately 1.5 hours, with short breaks provided between sessions to maintain engagement. 

\begin{wrapfigure}{l}{0.35\textwidth}
\includegraphics[width=\linewidth]{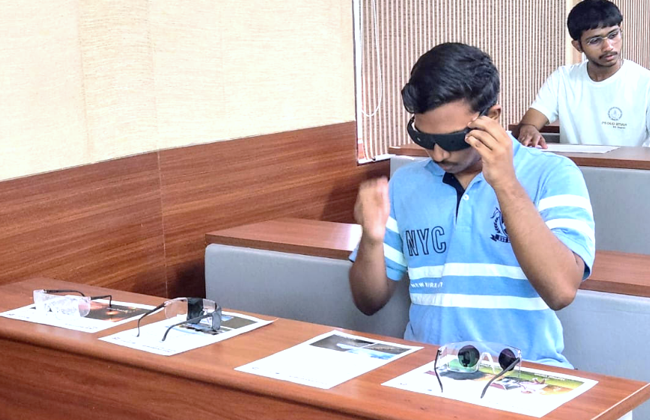}  
\caption{A Participant exploring a Visual impairment simulation}
\label{fig:simulaion} \vspace{-0.75cm}
\end{wrapfigure}

Additionally, four thematic booths were available throughout the day, offering hands-on exploration across key areas of accessibility. These included demonstrations of assistive technologies using both accessible and non-accessible examples and disability simulations as shown in Fig. \ref{fig:simulaion}, case studies highlighting real-world scenarios such as recent accessibility-related lawsuits, automated testing tool walkthroughs using platforms like WAVE and AXE, and lightning tech talks on emerging topics such as AI for accessibility. Facilitated by volunteers, these booths provided participants with the flexibility to engage with practical applications at their own pace during breaks. The workshop concluded with a post-survey and a structured reflection on lessons learned. \vspace{-0.55cm}

\subsubsection{Topics Covered in the Workshop}
To achieve the intended learning outcomes, we developed a structured course consisting of five instructional modules. The course draws heavily from the W3C Web Accessibility Initiative’s curricula on web accessibility\footnote{\url{https://www.w3.org/WAI/curricula/}}, as well as insights from prior research on effective approaches to teaching accessibility~\cite{palan_teaching_2017,parthasarathy_teaching_2024-1}. We outline each module below\footnote{The detailed handout has been hidden for reviews, and will be made available as a link post reviews.}: \vspace{-0.55cm}

\subsubsection{Accessibility Basics Module:}
This module provides a foundational understanding of accessibility by situating it within broader conversations on inclusion, equity, and user experience. The emphasis is on moving beyond a compliance-driven mindset to recognizing the value of accessibility in enhancing digital experiences for people with disabilities (PWDs). It covers the ethical, legal, social, and economic rationale for accessibility, alongside introductory implementation principles. Key topics include the types of disabilities, common challenges faced by users, and the role of assistive technologies; global disability demographics and their implications for inclusive technology; international frameworks such as the UN Convention on the Rights of Persons with Disabilities (UNCRPD); and ICT-related legal mandates including India’s Rights of Persons with Disabilities Act, the U.S. Section 508, and the EU’s EN 301 549. The module also addresses the business case for accessibility, highlighting benefits such as market expansion, brand reputation, innovation, and risk mitigation. Foundational web accessibility principles are introduced through the POUR framework (Perceivable, Operable, Understandable, Robust) and the structure of the WCAG guidelines. Finally, strategies for organizational adoption—such as leadership commitment, internal policy, and process integration—are discussed. \vspace{-0.55cm}

\subsubsection{Accessibility Design Module:}
This module introduces the concept of universal design and reinforces that accessibility is a shared responsibility extending beyond engineers to include designers, product managers, and other stakeholders. It promotes interdisciplinary collaboration—particularly between developers and UX designers—to ensure inclusivity from the outset. The session explores the distinction between individual accommodations and universal design approaches, applying core universal design principles to digital product development. WCAG is examined from a designer’s perspective, focusing on visual design, information design, navigation, and interaction. Topics like accessible media elements such as alt text and form feedback, animation considerations, and error handling were covered. The module concludes with guidance on preparing accessibility annotations for developer handoffs. \vspace{-0.55cm}

\subsubsection{Accessibility Development Module:}
Focusing on the technical side of building accessible applications, this module covers standards-based development practices and the use of semantic HTML and ARIA (Accessible Rich Internet Applications) to support assistive technologies. It explores how to meet WCAG success criteria through development techniques such as using standard controls, building custom components that are accessible, and creating accessible layouts, tables, dynamic content, and forms. Participants are introduced to methods for debugging and testing accessibility, including reproducing issues, performing quality assurance using automated and manual tools, and testing with screen readers and other assistive technologies. \vspace{-0.55cm}

\subsubsection{Accessibility Reporting Module:}
This module addresses how to identify, document, and remediate accessibility issues within software projects. It covers techniques for assessing the severity and user impact of accessibility bugs, categorized by harm potential, task blockage, or usability disruption. Participants are introduced to standardized reporting formats, such as the Accessibility Conformance Report (ACR) and the Voluntary Product Accessibility Template (VPAT). In addition, the session introduces practical accessibility testing techniques—including both automated tools and manual methods—as a precursor to effective reporting. The module also explores remediation strategies, including evaluating the feasibility and scalability of fixes, deciding when to refactor versus redesign, and effectively communicating issues to development teams. \vspace{-0.55cm}

\subsubsection{Teaching Accessibility Module:}
Recognizing the importance of embedding accessibility into CS education, this module provides strategies and resources for educators. Drawing on current research and leading teaching initiatives, the session introduces pedagogical frameworks such as active learning, empathy-based exercises, and inclusive design thinking~\cite{baker_systematic_2020,lewthwaite_accessible_2020}. It discusses how to integrate accessibility into core computing courses like web development, human-computer interaction, and software engineering. Participants are introduced to Oleson et al.’s open-access textbook Teaching Accessibility, which includes ready-to-use assignments, case studies, and WCAG-aligned content~\cite{oleson_teaching_2024}. Finally, the session highlights the value of community-based learning models such as Communities of Practice (CoPs) for fostering sustained engagement among educators and practitioners~\cite{parthasarathy_teaching_2024}. \vspace{-0.55cm}

\subsubsection{Instruments} 
To evaluate the effectiveness of the workshop, a pre- and post-workshop survey was conducted with all participants. The surveys were designed to assess changes in participants’ knowledge, attitudes, and self-reported confidence related to digital accessibility. The pre-survey, conducted during the workshop registration, captured baseline data on participants' familiarity with accessibility concepts and their background details. The post-survey, conducted at the end of the workshop, included questions to measure learning gains, along with additional items to capture participants' perceptions of the workshop content, delivery, and applicability to their professional contexts. The surveys combined Likert-scale items and open-ended questions to provide a mix of quantitative and qualitative insights into the workshop’s impact. \vspace{-0.35cm}

\section{Findings}
\label{sec:StudyResults}

\begin{figure}[h]
    \centering
    \includegraphics[width=\textwidth]{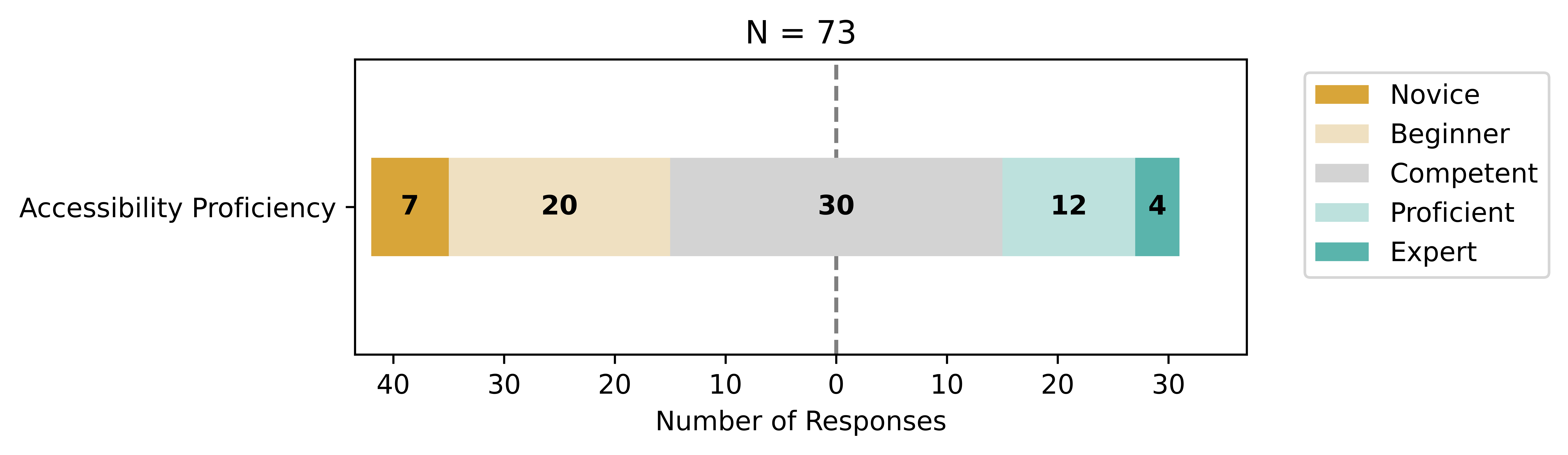}
    \caption{Accessibility Proficiency of Participants}
    \label{fig:accessibilityProficiency}
\end{figure}

The workshop was attended by 73 participants, comprising 33 professional software engineers, 28 educators, and 12 user experience (UX) designers. The group represented a wide range of professional experience: 17 participants had 0–2 years, 19 had 3–5 years, 18 had 6–10 years, and 19 had over 10 years of experience. Notably, 4 participants self-identified as having a disability, including two with visual impairments, one with a cognitive disability, and one with a speech-related condition. While 16 participants had prior exposure to accessibility—primarily through organizational training programs—the majority had no previous formal experience with accessibility concepts or practices. Fig ~\ref{fig:accessibilityProficiency} shows the self-reported accessibility proficiency levels of the participants. Interestingly, all four participants who identified as \textsc{Experts} in accessibility had over 10 years of professional experience, and two of them reported having a disability. All four experts were from the software industry. Additionally, among the 12 participants who rated themselves as \textsc{Proficient}, only two were educators, highlighting a notable disparity in accessibility expertise between academia and industry.

\begin{figure}[]
    \centering
    \includegraphics[width=0.65\textwidth]{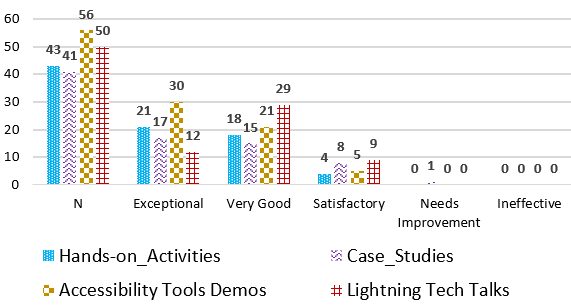}
    \caption{Activities Feedback from participants}
    \label{fig:activityFeedback} \vspace{-0.35cm}
\end{figure}

%\subsection{Effectiveness of Experiential Activities}
The feedback chart (Fig \ref{fig:activityFeedback}) reveals that all four activities of the workshop, the \textit{Hands-on Activities}, \textit{Case Studies}, \textit{Accessibility Tools Demos}, and \textit{Lightning Tech Talks}, were received positively by participants. When participants were asked to rate the usefulness of the activities, the majority marked them as either \emph{Exceptional} or \textit{Very Good}, with Accessibility Tools Demos receiving the highest number of \textit{Exceptional} ratings (30), followed by hands-on session on assistive technologies (21). Only a small number rated any session as \textit{Satisfactory}, and just one participant marked \textit{Needs Improvement} for Case Studies, with no activity labeled as \textit{Ineffective}. This suggests that the workshop's diverse pedagogical formats effectively engaged participants and were seen as valuable, with the accessibility tools demos and lightning tech talks standing out as particular highlights.

\begin{figure}[]
    \centering
    \includegraphics[width=0.95\textwidth]{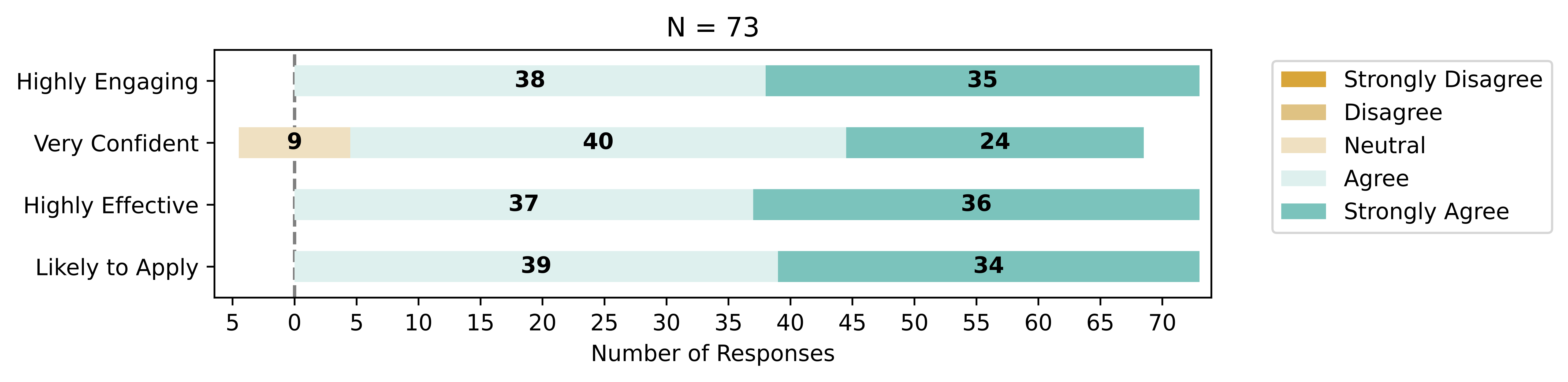}
    \caption{Post workshop Feedback from participants (N=73)} 
    \label{fig:postWorkshop} \vspace{-0.35cm}
\end{figure}

The post-workshop feedback from 73 respondents reflects a highly positive perception of the workshop’s design and impact (Fig. \ref{fig:postWorkshop}). A total of 73 participants rated their experience across four dimensions: engagement, confidence, effectiveness, and likelihood of applying the learning.

On the engagement aspect, 38 participants agreed and 35 strongly agreed that the workshop was \textit{Highly engaging}. Regarding confidence, while 9 participants remained neutral, 40 agreed and 24 strongly agreed that they felt \textit{very confident} after the workshop. In terms of overall effectiveness, 37 participants agreed and 36 strongly agreed that the workshop was \textit{highly effective}. Similarly, 39 participants agreed and 34 strongly agreed that they were likely to apply what they learned in the near future at their workplace, highlighting the practical relevance of the sessions. Additionally, 52\% of participants reported a significant shift in their perception of accessibility and their personal responsibility in promoting it, while the remaining participants noted a moderate change in attitude. When asked whether they would recommend the workshop to their personal and professional networks, all participants (100\%) responded affirmatively.

We also included an open-ended question in the post-workshop survey, inviting participants to suggest improvements and propose topics for future workshops. Their responses were analyzed using thematic analysis, following the approach outlined by Saldana~\cite{saldana_coding_2021}. Five themes emerged out of this analysis, which are described below: 

\textit{(1) Advanced Practices in Inclusive Design and Development}: Participants expressed strong interest in deepening their skills in advanced accessibility techniques, particularly related to modern UI/UX practices. Topics such as using accessible design tokens, customizing components, creating accessible user flows, and developing semantic HTML and ARIA-based code emerged as key areas for further exploration. Additionally, several responses (12) highlighted the need for better design tools and Figma plugins that support accessibility compliance from the early stages of development. 

\textit{(2) AI and Emerging Technologies in Accessibility}: A significant number (27) of participants were curious about the intersection of AI and accessibility. Requests included understanding the role of AI chatbots, evaluating the accuracy of AI-generated alt text, and exploring how machine learning could enable personalized accessibility features. Participants also expressed a desire to learn how virtual assistants and automated WCAG testing tools can be leveraged to improve user experiences for people with disabilities.

\textit{(3) Inclusive Design for Diverse Contexts and Users}: Another strong theme was designing for diversity. 13 Participants called for focused learning on accessibility for neurodiverse users, low-vision users, and the needs of learners in online and remote settings. Notably, this interest was more pronounced among educators than among technology professionals. 

\textit{(4) Organizational Strategies and Policy Dimensions}:
Several responses (35) emphasized the importance of organizational commitment and legal awareness. There was notable interest in building company-wide accessibility initiatives, understanding public policy’s role in shaping digital accessibility, and staying informed about compliance risks and legal trends. Participants were keen to understand how to move beyond checklist-driven approaches and embed accessibility into the broader culture of product development.

\textit{(5) Need for Practice-Based Learning and Real-World Examples}: Finally, participants advocated for more experiential and case-based learning in future sessions. They expressed interest in seeing real-world examples of accessibility successes and failures, particularly through detailed case studies. 19 participants also wanted guidance on conducting user research with people with disabilities and on specific challenges such as making documents, charts, and graphs more accessible.
\section{Discussion}
\label{sec:takeaway}

The workshop provided a valuable opportunity to engage computer science educators and professionals in hands-on, practice-oriented accessibility training—an area often underrepresented in formal computing education in India. Insights drawn from both quantitative feedback and qualitative reflections highlight what worked well, the challenges encountered, and key considerations for scaling similar initiatives in the future.

\subsection{What Worked Well}
The workshop was positively received across participant groups, with feedback reflecting strong engagement, perceived effectiveness, and practical relevance. The diverse participant pool—including educators, software engineers, and UX designers—enabled a multidisciplinary dialogue on accessibility, which many noted as enriching. The structured sessions, supported by hands-on components and thematic booths, were seen as effective in contextualizing abstract concepts and introducing real-world tools and techniques.

All four core activities—Hands-on Activities, Case Studies, Accessibility Tools Demos, and Lightning Tech Talks—were rated as either Exceptional or Very Good by the majority of participants, with the Accessibility Tools Demos receiving the highest number of Exceptional ratings. The lightning tech talks and assistive technology sessions were also specifically appreciated for their practical insights and exposure to emerging trends.

The post-workshop survey revealed that 100\% of participants were willing to recommend the workshop to peers, and 90\% agreed or strongly agreed that it was highly effective and engaging. Additionally, 52\% reported a significant shift in their perception of accessibility and their responsibility in advancing it, suggesting that the experiential design of the workshop was successful in promoting attitudinal change.

Thematic analysis of open-ended responses further supported this, with participants expressing strong interest in deeper, practice-oriented learning across areas such as advanced UI design, AI applications, and organizational accessibility strategies. These findings underscore the workshop’s success in both raising awareness and sparking professional curiosity.

\subsection{What Could Have Been Better}
Despite its overall success, the workshop also revealed several areas for improvement. While the hands-on activities were appreciated, a few participants, especially those with limited technical backgrounds, experienced difficulty engaging with certain tools and development-focused sessions. For instance, Session 3 on Accessible Software Engineering, which included complex technical content, was perceived as challenging by some educators and designers, who felt less equipped to follow the material. This indicates a need for differentiated instructional strategies or scaffolding based on participant roles and prior experience.

Another noted limitation was the reliance on simulations for demonstrating assistive technologies as shown in Fig. \ref{fig:simulaion}. While these exercises were effective in our workshop, simulations have been problematic as an educational tool, as suggested by several prior studies, including~\cite{tigwell_nuanced_2021} and~\cite{bennett_promise_2019} and is advisable to replace them by integrating lived experiences into the instructional design.

Additionally, some participants expressed a desire for more domain-specific examples, particularly in education, mobile UX, and neurodiversity-focused design. Educators, in particular, sought examples relevant to online learning platforms and inclusive pedagogy. Participants also highlighted the value of supplementary materials—such as toolkits, curated readings, or video tutorials—to support continued learning beyond the one-day format.

\subsection{Recommendations for Future Efforts}

Based on participant feedback and our observations, we recommend the following to enhance the effectiveness of future workshops:\\
\textit{(1) Distribute Pre-Workshop Materials:} Sharing tutorials, reading resources, and tool guides ahead of the session can better prepare participants and ensure smoother engagement with hands-on activities.\\
\textit{(2) Customize Content by Role or Domain:} Including domain-specific examples (e.g., education, mobile apps, government services) and role-based tracks (e.g., educators, developers, designers) can increase relevance and applicability.\\
\textit{(3) Include Case Studies and Practical Use Cases:} Real-world examples, including both successful and problematic implementations, can enhance conceptual understanding and foster critical reflection.\\
\textit{(4) Encourage Peer Learning and Networking:} Structured opportunities for participant interaction—such as discussion groups or experience-sharing booths—can deepen learning and build a sense of community.\\
\textit{(5) Provide Opportunities for Continued Engagement:} Offering follow-up sessions, access to shared resources, or an online platform for continued dialogue can help sustain momentum beyond the workshop.

%\subsection{Broader Implications}

%This workshop demonstrates the value of embedding accessibility into mainstream professional development and computing education in India. By emphasizing experiential, hands-on learning and fostering dialogue across academia and industry, such initiatives can play a crucial role in equipping professionals and educators with the mindset and skills needed to build more inclusive technologies. As digital systems increasingly shape education, work, and public services, integrating accessibility from the ground up is not just a technical requirement—it is an ethical imperative. Scaling and localizing these efforts can contribute meaningfully to national and institutional goals for digital inclusion.
\section{Conclusion}
\label{sec:conc}

We designed and delivered an informal accessibility education intervention in the form of a hands-on workshop. The workshop underscores the need to embed accessibility within computing education and professional training in India. Participants from academia and industry reported increased confidence, improved practical understanding, and a meaningful shift in their attitudes toward accessibility. The outcomes highlight the effectiveness of experiential learning in promoting inclusive design mindsets and bridging knowledge gaps across sectors. As digital technologies become increasingly central to everyday life, scaling such efforts is critical to advancing digital inclusion and creating a more accessible future. To sustain impact, such interventions must be continued and followed up to assess whether participants apply the accessibility learnings within their workplace contexts.

\bibliographystyle{IEEEtran} % to avoid the duplication of URLs in the references 
\bibliography{references}

\end{document}